\documentclass[
aps,
reprint,
prr,
superscriptaddress,
nobibnotes,
 amsmath,
 amssymb,
]{revtex4-2}

\usepackage{graphicx}
\usepackage{dcolumn}
\usepackage{bm}
\usepackage{hyperref}
\usepackage{cleveref}
\usepackage{braket}
\usepackage{tikz}
\usetikzlibrary{positioning}
\usepackage{xcolor}
\usepackage{algorithm}
\usepackage{algpseudocode}
\usepackage{bbold}
\usepackage{setspace}

\tikzset{node1/.style={circle, draw=mypink, fill=mypink!10, thick}}
\tikzset{node2/.style={circle, draw=myblue, fill=myblue!10, thick}}
\tikzset{myarrow/.style={->, thick}}

\newcommand{\IIZ}[1]{{\color{black}#1}}
\newcommand{\JC}[1]{{\color{black}#1}}

\begin{document}

\title{Robust microwave cavity control for NV ensemble manipulation}

\author{Iñaki Iriarte-Zendoia}
\author{Carlos Munuera-Javaloy}
\affiliation{Department of Physical Chemistry, University of the Basque Country UPV/EHU, Apartado 644, 48080 Bilbao, Spain}
\affiliation{EHU Quantum Center, University of the Basque Country UPV/EHU, Leioa, Spain}
\author{Jorge Casanova}
\affiliation{Department of Physical Chemistry, University of the Basque Country UPV/EHU, Apartado 644, 48080 Bilbao, Spain}
\affiliation{EHU Quantum Center, University of the Basque Country UPV/EHU, Leioa, Spain}

\begin{abstract}
Nitrogen-vacancy (NV) center ensembles have the potential to improve a wide range of applications, including nuclear magnetic resonance spectroscopy at the microscale and nanoscale, wide-field magnetometry, and hyperpolarization of nuclear spins via the transfer of optically induced NV polarization to nearby nuclear spin clusters. These NV ensembles can be coherently manipulated with microwave cavities, that deliver strong and homogeneous drivings over large volumes. However, the pulse shaping for microwave cavities presents the challenge that the external controls and intra-cavity field amplitudes are not identical, leading to adverse effects on the accuracy of operations on the NV ensemble. In this work, we introduce a method based on Gradient Ascent Pulse Engineering (GRAPE) to optimize external controls, resulting in robust pulses within the cavity while minimizing the effects of cavity ringings. The effectiveness of the method is demonstrated by designing both $\pi$ and $\pi/2$ pulses. These optimized controls are then integrated into a PulsePol sequence, where numerical simulations reveal a resilience to detunings five times larger than those tolerated by the sequence constructed using standard controls.
\end{abstract}

\maketitle

\section{\label{sec: intro} Introduction}

The Nitrogen-vacancy (NV) center~\cite{NVRev_Wrachtrup, QTech_Awschalom} is a promising platform for quantum applications owing to its ease of control, which includes optical initialization and readout, microwave coherent manipulation, and long coherence times even at room temperature. NV ensembles are able to detect signals from extremely small volumes, making them promising candidates for spectroscopy at micro- and nanoscale levels~\cite{microscale_Glenn, microscale_Arunkumar, microfluids_Bucher, MicroNanoscale_Allert, nanoscale_Mamin, nanoscale_Staudacher, nanoscale_Segawa, nanoscale_Schirhagl, nanowide_Hong, JInsect_Alsina, HighField_Munuera, AERIS_Munuera}. NV ensembles are also being researched for wide-field magnetometry~\cite{nanowide_Hong, livingCells_LeSage, widefield_Sengottuvel, widefield_Guo}, and for dynamical nuclear polarization~\cite{multispin_Pagliero, orientationIndependent_Ajoy, DNP_Tetienne, PulsePol_Schwartz, HP-C13_Chen}. In the latter case, the polarization of nearby nuclear spin clusters benefits from the large number of NV centers within the ensemble, as each defect transfers a fraction of polarization to the nuclei.

\JC{NV ensembles require strong and uniform drivings to perform accurate unitary operations across the entire ensemble. To achieve such fields, diamonds are often placed within a resonant microwave antenna, which can be  modeled as a cavity in the region containing the NV ensemble (see section CAVITIES in the Supplementary Material~\cite{SupplementaryMaterial})}.  \JC{In this framework, under an external control, the amplitude of the intra-cavity field gradually converges to a value proportional to the applied control, at a rate determined by the cavity ringing factor~\cite{SupplementaryMaterial} (subsection Ringing).
The cavity must be engineered to provide the necessary driving strength and spatial homogeneity across the NV ensemble area~\cite{CohMan_Yudilevich, QLogic_Arunkumar, OptimPlanar_Opaluch, 3D_Kapitanova, Circular_Yaroshenko}, to enable fast and uniform operations on the defects (the same principle would apply in other contexts, such as molecular polarizers in NMR~\cite{Macroscopic_Marshall}).
Note that this occurs in cavities with low ringing factors~\cite{OptimPlanar_Opaluch, Lasers_Siegman, MW_Pozar}, making this regime particularly interesting for applications with large NV ensembles, which is our primary focus.
}

\JC{To introduce our method, we firstly provide an overview of the distinct techniques employed as a function of the ringing factor, with detailed explanations available in the Supplemental Material in Ref.~\cite{SupplementaryMaterial} (subsection Control algorithms).} \JC{For large values of the ringing factor, the cavity reaches its steady state almost instantly, and the shape of the intra-cavity amplitude matches that of the external controls up to a normalization factor (see Eq.~(\ref{eq: ODE}) and~\cite{SupplementaryMaterial})}. \JC{In this regime, standard algorithms such as the Gradient Ascent Pulse Engineering (GRAPE)~\cite{GRAPE_Khaneja} or techniques based on a superoperator description, such as the Universally Robust Quantum Control (URC)~\cite{URC_Poggi} method, are adequate for pulse shaping optimization leading to error resilient controls~\cite{Robust_Said}}. \JC{For intermediate values of the ringing factor, where external controls and intra-cavity amplitudes slightly differ, pulse design is still achievable via a piecewise linear approximation~\cite{piecewise_Rasulov}.}

\JC{However, when the ringing factor is small, the cavity's response time becomes significant, leading, among other effects, to detrimental ringing~\cite{SupplementaryMaterial}. Additionally, the intra-cavity amplitudes deviate substantially from the external controls, which makes previous approaches ineffective.}

\JC{While GRAPE has been used to tune external controls in distinct setups~\cite{TransferFunction_Motzoi}, we have designed a quantum control algorithm for the case of NV ensembles driven by cavity fields. In particular, our method optimizes external controls to generate robust intra-cavity pulses that maintain performance across an NV ensemble (i.e., effectively operating over large detuning errors). This remains true even in the low ringing factor regime, where suppressing ringing effects is crucial.} We employ our algorithm to generate robust $\pi$ and $\pi/2$ pulses, which are then incorporated into a PulsePol~\cite{PulsePol_Schwartz} sequence capable of operating over a wide detuning range.

In Section~\ref{sec: system}, we provide the theoretical details of the system under study. Section~\ref{sec: algorithm} outlines the algorithm. In Section~\ref{sec: results}, we apply the algorithm to enhance the robustness of the PulsePol building blocks against detunings, aiming to improve its inherent robustness. Finally, in Section~\ref{sec: discussion}, we discuss our findings and explore possibilities for future research.

\begin{figure*}
	\includegraphics[width=\textwidth]{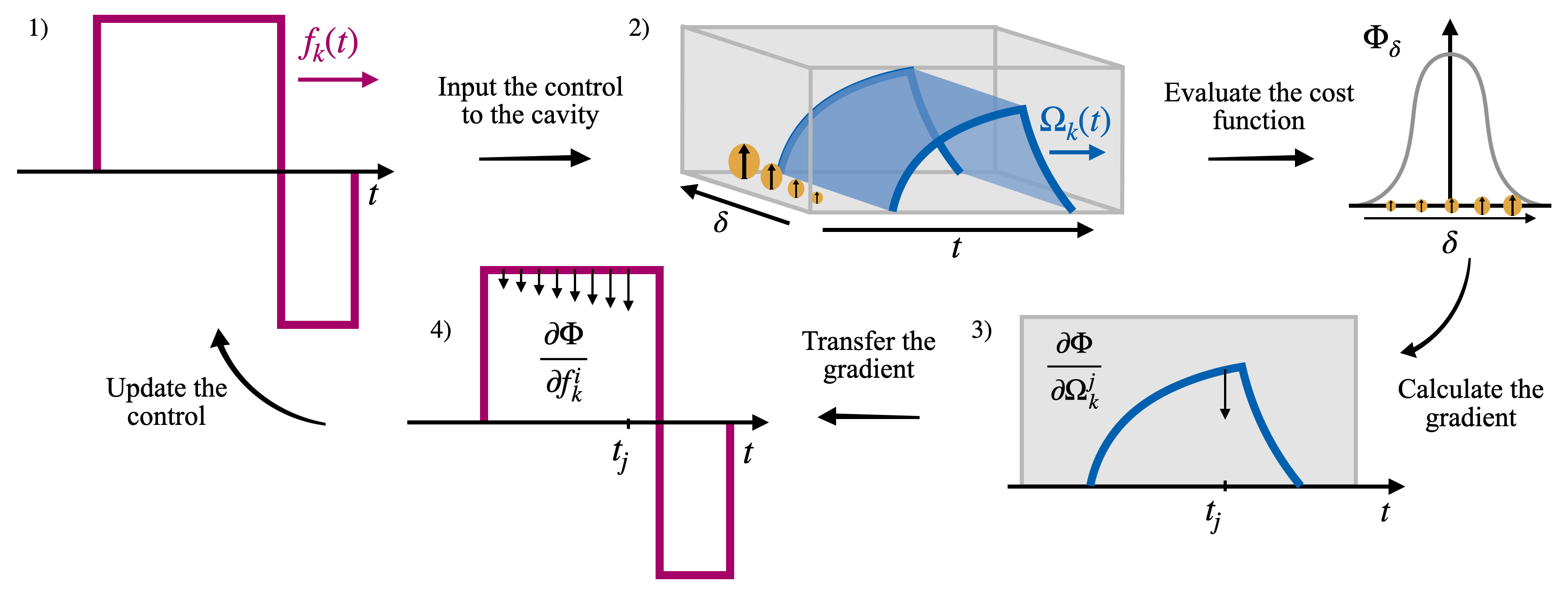}
	\caption{1) The external control (pink) is inputted to the ODE~(\ref{eq: ODE}) to obtain $\Omega_k(t)$. 2) The internal pulse is applied over an ensemble of spins, i.e. copies of~(\ref{eq: optimisation_hamiltonian}) with different values of $\delta$, represented through yellow balls with different sizes. This yields unitaries with varying fidelities \IIZ{$\Phi_\delta$}. 3) The gradient of each fidelity with respect to $\Omega_k^j$ is computed and summed over the detunings. 4) Subsequently, the gradients are transferred to the external control to update it. Since the ODE makes the value of $\Omega_k(t)$ dependent on the external control value at previous times, transferring the gradient corresponding to $\Omega_k(t_j)$ affects to the gradients applied to $f_k(t_i)$ for $t_i\leq t_j$. }
	\label{fig: diagram}
\end{figure*}

\section{\label{sec: system} The system}
We consider the Hamiltonian of a spin-$\frac{1}{2}$ Hamiltonian describing a generic NV in an ensemble under the influence of MW drivings~\cite{SupplementaryMaterial}:
\begin{equation}
	H(t) = \frac{\delta}{2}\sigma_z + \sum_{k\in\{x,y\}} \frac{\Omega_k(t)}{2}\sigma_k.
	\label{eq: optimisation_hamiltonian}
\end{equation}
Here, $\sigma_{x,y,z}$ are Pauli operators, $\delta$ is the detuning of the MW field relative to the NV resonance frequency, which varies from one NV to another in the ensemble; and $\Omega_x(t)$ and $\Omega_y(t)$ denote the time dependent amplitudes of the applied MW pulses. When the spin is located inside a resonant cavity, the intra-cavity amplitudes  ($\Omega_x(t)$ and $\Omega_y(t)$) are generated by  external controls ($f_x(t)$ and $f_y(t)$) through the following ordinary differential equation (ODE)~\cite{Thesis_Tratzmiller, PhaseNoise_Bienfang, SupplementaryMaterial, DynamicResponse_Lawrence}:
\begin{equation}
	\dot{\Omega}_k(t) = \gamma\Omega_\mathrm{max}f_k(t) - \gamma \Omega_k(t)\,, \; k\in\{x,y\}.
	\label{eq: ODE}
\end{equation}
Here, $f_k(t)\in[-1,1]$ represents the normalized amplitude of the external control, $\gamma$ is the ringing factor and $\Omega_\mathrm{max}$ denotes the maximum intra-cavity amplitude. Solving Eq.~(\ref{eq: ODE}) yields:
\begin{equation}
	\begin{aligned}
		\Omega_k(t) &= e^{-\gamma (t-t_0)} \Omega_k(t_0) + \\
		&+ e^{-\gamma (t-t_0)} \Omega_\text{max}\int_{t_0}^{t}\gamma f_k(t^\prime)e^{\gamma (t^\prime-t_0)} dt^\prime,
	\end{aligned}
    \label{eq: ODE_solution}
\end{equation}
where we consider $\Omega_k(t)=0$ at $t=t_0$.

To provide intuition for the dynamics of $\Omega_k(t)$, consider the cavity's response to an external control. If the cavity is initially driven by a constant control such that $f_k(t)=f_0$, the amplitude of $\Omega_k(t)$ will asymptotically approach a steady-state value of $f_0\Omega_\text{max}$, i.e., $\Omega_k(t\rightarrow \infty)=f_0\Omega_\text{max}$. When the external control is turned off (i.e., $f_k$ is set to zero), $\Omega_k(t)$ decays exponentially over time, a phenomenon known as cavity ringing. Minimizing this ringing is crucial to prevent pulse overlap in sequences with multiple nested pulses.

\section{\label{sec: algorithm} The algorithm}

\begin{table}
	\begin{algorithm}[H]
	\caption{\raggedright{\small Chain-GRAPE (our adaptation for cavities). }}
		\begin{algorithmic}[1]
			\Require \parbox[t]{190pt}{\setstretch{1.1}\raggedright Initial ansatz: $f_x$ and $f_y$. \\
				Vector of detunings: $\vec\delta$. \\
				Weight of detuning $\delta$: $w_\delta$. \\
				Ringing correction: $\alpha$. \\
				Threshold: $\epsilon \ll 1$.
				}
			\vspace{3pt}
			\Ensure \parbox[t]{190pt}{\setstretch{1.1}\raggedright Optimized controls: $f_x$ and $f_y$. \\
				}
			\State Calculate $\Omega_k^{1\times N}$ with $f_k^{1\times n_f}$.
			\While{$1-\Phi \geq \epsilon$}
				\ForAll{$\delta \in \vec\delta$}
					\State Calculate $U_j$ from $H(\delta, \Omega_x^j, \Omega_y^j)$ for all $j\leq N$.
					\State Calculate $\partial \IIZ{\Phi_\delta} / \partial \Omega_k^j$.
					\State $D_k(j) \gets D_k(j) + w_\delta \cdot \partial\IIZ{\Phi_\delta} / \partial \Omega_k^j$ for all $j\leq N$. (This trains a $\delta$-dependent behavior, providing robustness.)
				\EndFor
				\State $D_k(N) \gets D_k(N) - \alpha\cdot \Omega_k^N$ (Correction for ringing.)
				\State $G_k(i) \gets G_k(i) + D_k(j) \cdot \partial \Omega_k^j/\partial f_k^i$ for all $i\leq\mathtt{ceil}(j/r)$ (Transfer the gradient to the external driving.)
				\State Update $f_{x(y)}$ with the gradients $G_{x(y)}$.
			\EndWhile
			\Statex\Comment{$D_k$ and $G_k$ account for $\frac{\partial \Phi}{\partial\Omega_k}$ and $\frac{\partial \Phi}{\partial f_k}$ respectively. They consist of a vector of $N$ and $n_f$ elements.}
		\end{algorithmic}
		\label{alg: 1}
	\end{algorithm}
\end{table}

We introduce an algorithm termed Chain-GRAPE, which generates external controls $f_k(t)$ that result into time-dependent intra-cavity amplitudes, $\Omega_k(t)$. These are capable of implementing arbitrary target unitaries --in the following $U_{F}$-- while remaining robust against detunings of several MHz. The discrete nature of the obtained $\Omega_k(t)$ results in a unitary of the kind $U(\delta)=U_N(\delta)\cdots U_2(\delta)\cdot U_1(\delta)$, with
\begin{equation}
    U_j(\delta) = \exp\left\{-2\pi i \left(\frac{\delta}{2}\sigma_z + \sum_k \Omega_k^j \frac{\sigma_k}{2} \right) dt\right\},
    \label{eq: unitary}
\end{equation}
where $dt=\frac{T}{N}$, $T$ refers to the total evolution time and $N$ the number of time steps. In Eq.~(\ref{eq: unitary}), we take $\Omega_k^j~\equiv~\Omega_k(jdt)$.

We discretize $\Omega_k(t)$ and $f_k(t)$ using different time steps, $dt$ and $\Delta t$ respectively, and relate their corresponding number of steps ($N$ and $n_f$) by $n_f \cdot r = N$, where the ratio  $r = \Delta t / dt$. This approach allows us to adapt the algorithm for a small number of time steps ($n_f$), which facilitates optimization, while a large $N$ results in an almost continuous curve for $\Omega_k(t)$, enabling an accurate simulation of the intra-cavity dynamics. Additionally, the discretization of  $f_k(t)$ is designed to be compatible with experimental devices that generate controls with a limited sampling rate.

We define the cost function ($\Phi$)~\cite{GRAPE_Khaneja} for our optimization problem  as
\IIZ{
\begin{equation}
    \Phi = \sum_\delta w_\delta \Phi_\delta,
    \label{eq: gate_fid}
\end{equation}
}
where $\braket{A|B} \equiv \mathrm{tr}(A^\dagger B)$ and $w_\delta$ is the weight assigned to \IIZ{$\Phi_\delta=|\braket{U(\delta)|U_F}|^2/4$}. The factor of $1/4$ arises from normalizing the trace. By assigning a larger weight to a certain detuning $\delta$, we guide the optimizer to prioritize the matching of $U(\delta)$ to $U_F$. In other words, this approach emphasizes the accuracy of the resulting unitary at particular detuning ranges.

The next step is to select an optimizer that successfully yields the controls that maximize the cost function. Among the wide range of optimizers, we focus on gradient-based methods, such as ADAM \cite{ADAM_Kingma}. This method uses the gradient of $\Phi$ with respect to certain parameters (in our case $f_k^i~\equiv~f_k(i\Delta t)$) guiding the algorithm toward a maximum of $\Phi$ in parameter space. To calculate the gradient of $\Phi$ with respect to $f_k^i$ we split it into two parts as follows:
\begin{equation}
	\frac{\partial\Phi}{\partial f_k^i} = \sum_{j=1}^N\frac{\partial\Phi}{\partial\Omega_k^j}\frac{\partial\Omega_k^j}{\partial f_k^i}.
	\label{eq: full_gradient}
\end{equation}
The right hand part is computed differentiating Eq.~(\ref{eq: ODE_solution}):
\begin{equation}
	\frac{\partial\Omega_k^j}{\partial f_k^i} = e^{-\gamma jdt}\Omega_\mathrm{max} \left( e^{\gamma i\Delta t} - e^{\gamma (i-1)\Delta t} \right),
	\label{eq: grad2}
\end{equation}
with $i=1,...,i_c$ and $i_c=\mathtt{ceil}(j/r)$, where $\mathtt{ceil}(x) = \lceil x \rceil = \mathrm{min}\{n\in \mathbb{Z} : x \leq n\}$~\cite{SupplementaryMaterial}. The left-hand term is computed following Ref.~\cite{GRAPE_Khaneja}:
\begin{equation}
	\begin{aligned}
    		\frac{\partial\Phi}{d \Omega_k^j} = & -\pi i  dt \cdot \\
		&\cdot \sum_\delta w_\delta \text{Re}\bigl( \braket{P_j(\delta) | \frac{\sigma_k}{2} X_j(\delta)}\braket{X_j(\delta)|P_j(\delta)} \bigr),
    \label{eq: gate_GRAPE}
    \end{aligned}
\end{equation}
where $P_j(\delta) = U_{j+1}^\dagger(\delta)\cdots U_N^\dagger(\delta) U_F$ and $X_j(\delta)=U_j(\delta)\cdots U_1(\delta)$. 

\begin{figure*}[t!]
	\includegraphics[width = \linewidth]{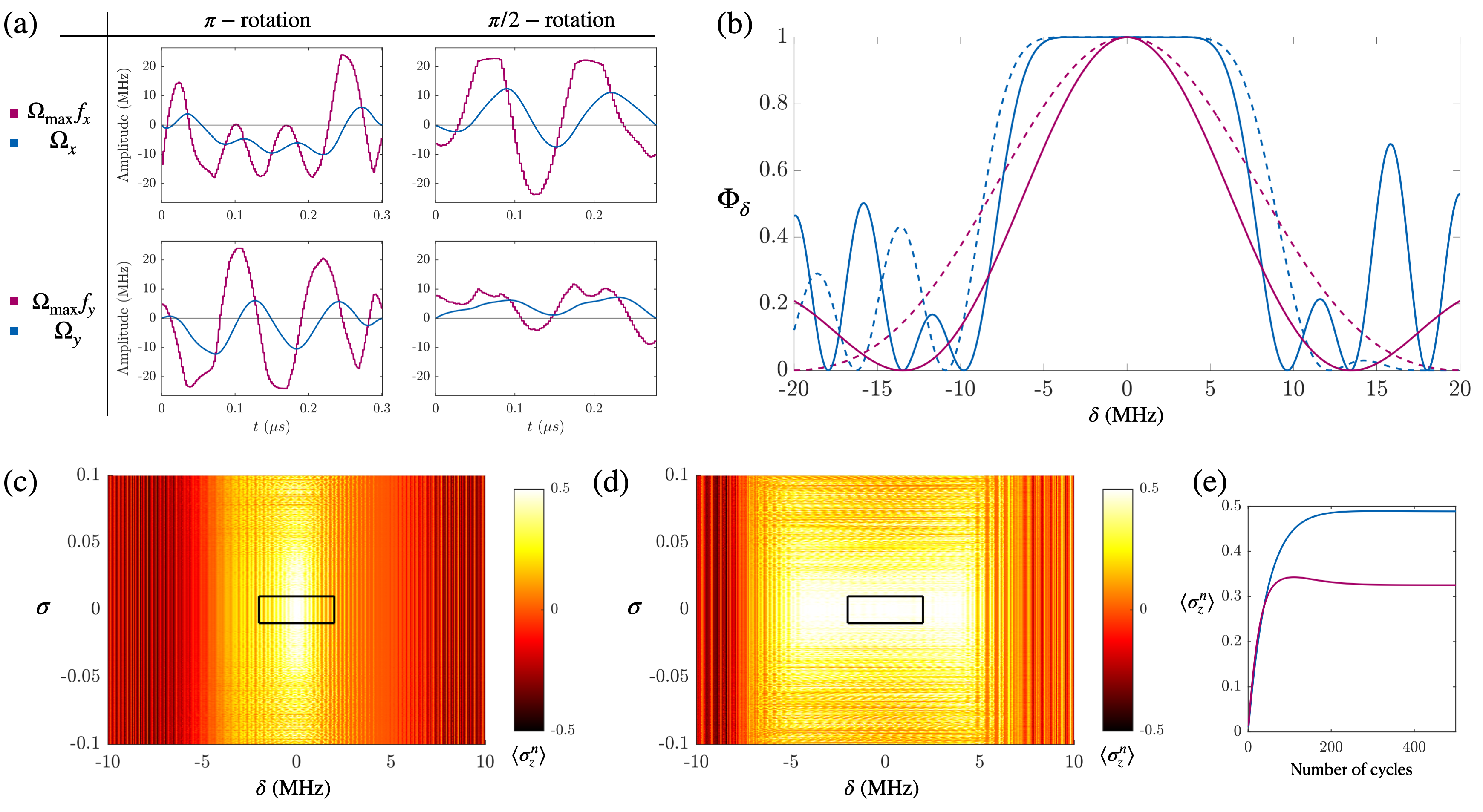}
	\caption{\textbf{(a)} Optimized external (pink) and internal (blue) microwave fields for $\pi$ and $\pi/2$ rotations along the X-axis. The blue curves, representing $\Omega_k^j$, end with zero amplitude ($\Omega_k^N \simeq 0$), effectively mitigating cavity ringing effects. \textbf{(b)} Solid lines represent the fidelity $|\braket{U(\delta)|U_F(\theta)}|^2$ for $\theta=\pi/2$, while dashed lines correspond to $\theta=\pi$. Remarkably, the blue curves generated with our optimized external controls show resilience to detunings up to 5 MHz. On the contrary,  pink curves obtained with the standard controls~\cite{SupplementaryMaterial, Thesis_Tratzmiller}, handle detunings of only up to 1 MHz. Figures \textbf{(c)} and \textbf{(d)} depict the nuclear polarization $\langle \sigma_z^n \rangle$ (see color-bar) as a function of errors $\delta$ and $\sigma$ after applying the protocol built with the standard pulses and our optimized pulses, respectively. Two PulsePol sequences were applied before reinitializing the NV completing one cycle, and 500 cycles were performed before measuring the nuclear polarization. In \textbf{(e)} we depict the average polarization curve inside the regions marked by the squares in figures (c) and (d). These regions are delimited by the constraints $|\delta| \leq 2\,$MHz and $|\sigma| \leq 0.01$. The curve achieved with our optimized external controls (blue) exceeds the one obtained with the standard approach (pink). The values of the parameters used in the simulations are $\gamma=20\,$MHz, $\Omega_\mathrm{max}= 24\,$MHz, $B_z=0.015\,$T,  $A_x=4\,$kHz and $A_z=3.7\,$kHz.}
	\label{fig: results}
\end{figure*}

As discussed in Section~\ref{sec: system}, ensuring that the field vanishes within the cavity at the end of the evolution is crucial. Failure to do so would result in undesired dynamics due to the residual amplitude of the controls.
To address this, we modify the $j=N$ element of Eq.~(\ref{eq: gate_GRAPE}) as
\begin{equation}
	\frac{\partial\Phi}{\partial\Omega_k^N} \rightarrow \frac{\partial\Phi}{\partial\Omega_k^N} -\alpha\Omega_k^N.
\end{equation}
This adjusts the cost function such that it is maximized when the intra-cavity amplitude at the last step, $\Omega_k^N$, is small.

This adjustment necessitates choosing an appropriate value for the proportionality constant $\alpha$ to balance the trade-off between maximizing (\ref{eq: gate_fid}) and suppressing ringing, enabling the algorithm to optimize both objectives simultaneously. The complete algorithm is explained in Alg.~\ref{alg: 1}~and Fig.~\ref{fig: diagram}.

\section{\label{sec: results}Results}
A key application of the proposed algorithm is in designing controls that generate robust pulses for specific pulse sequences, enabling their deployment across large NV ensembles.
For instance, in the case of the PulsePol~\cite{PulsePol_Schwartz} sequence (which is designed for polarization transfer), incorporating our method would further enhance its resilience to detuning errors~\cite{PulsePol_Schwartz, SymmetryBased_Sabba}, thereby extending its applicability to the many NVs scenario.

Now we demonstrate polarization transfer from an NV to a $^{13}$C nucleus in a wide range of detunings using our method. \IIZ{The Hamiltonian describing the system (for details see~\cite{SupplementaryMaterial}) consists of the sum of (\ref{eq: optimisation_hamiltonian}), a Larmor precession term generated by the magnetic field ($B_z$), and the perpendicular ($A_x$) and parallel ($A_z$) components of the coupling between the NV and nucleus.} To achieve gradual polarization transfer, we concatenate polarization cycles by applying the PulsePol sequence twice (note the delivered PulsePol uses our pulses as building blocks)  followed by NV reinitialization.

Our algorithm outputs external controls (see Fig.~\ref{fig: results}~(a) which result in a unitary that approaches
\begin{equation}
	U_F(\theta) = \exp(-i \theta \sigma_x/2),
	\label{eq: U_F}
\end{equation}
for $\theta=\pi$ and $\theta=\pi/2$ ($\pi$ and $\pi/2$ pulses respectively), covering detunings up to 5 MHz. These operations are five times more robust than the unitaries generated with the standard controls (see Fig.~\ref{fig: results}~(b) and Refs.~\cite{SupplementaryMaterial, Thesis_Tratzmiller} for a description of the standard controls). Since PulsePol requires unitaries along the Y-axis as well (i.e., we also have to target $U_F(\theta) = \exp(-i \theta \sigma_y/2)$), we need the controls ($f_x^\prime$ and $f_y^\prime $) such that  $f_x^\prime = -f_y$ and $f_y^\prime = f_x$~\cite{SupplementaryMaterial}.

To investigate the added robustness of PulsePol provided by the optimized pulses, we examined the loss of polarization due to detuning $\delta$, as well as to control errors in the Rabi frequency. We have modeled the latter as Gaussian noise, leading to the following modified controls: $\tilde f_k(t) =  f_k(t)\cdot[1+\sigma  \epsilon(t)]$, where $\epsilon(t)$ stands for random numbers following a unit normal distribution, and $\sigma$ denotes the deviation. Comparing Figs.~\ref{fig: results}~(c) and \ref{fig: results}~(d), we observe that the sequence generated with our optimized pulses exhibits significantly greater resilience to detuning errors compared to the standard approach. We note that detunings are the primary source of error in NV ensembles, while control amplitude deviations are typically around 1$\%$, highlighting the suitability of our method.

In Fig.~\ref{fig: results}~(e) we plot the average polarization curve, as it is a representative magnitude to compare the speed of the polarization transfer. For that, we first calculate the nuclear polarization after each cycle for each pair values ($\delta$, $\sigma$) enclosed within the black squares in Figs.~\ref{fig: results}~(c) and \ref{fig: results}~(d). The plotted curve is obtained taking the average over all the curves \IIZ{(for a more detailed analysis of robustness, see~\cite{SupplementaryMaterial})}. The results obtained using our pulses show higher polarization, see Fig.\ref{fig: results}~(e), confirming the enhanced robustness provided by our optimized controls.

\section{\label{sec: discussion}Discussion}
We have \IIZ{modified} an algorithm to optimize the external controls of a microwave resonant cavity, ensuring that the resulting intra-cavity amplitudes perform the desired unitary operation on NV centers across a broad range of detunings. By optimizing external controls to perform $\pi$ and $\pi/2$ pulses for NV ensembles, we have demonstrated that our algorithm generates robust controls that effectively mitigate ringing effects. Moreover, we have utilized these pulses to construct a PulsePol sequence with enhanced robustness, demonstrating their effectiveness in polarization transfer protocols. This improvement can be extended beyond NV-based methods to other hyperpolarization techniques such as DNP, PHIP, and SABRE \cite{DNP-PHIP-SABRE_Kovtunov, PHIP_Hovener, PHIP/DNP_Korzeczek}. Lastly, exploring the algorithm's applicability across various control techniques would be highly valuable. This includes potential integration with mechanical resonators \cite{mechanical_MacQuarrie}, sensing experiments involving ensembles controlled by microwave antennas \cite{QLogic_Arunkumar}, and superconducting circuits placed within microwave resonant cavities \cite{SC_Naghiloo}.

\section{Acknowledgements.-- } I.I.Z. acknowledges support from UPV/EHU Ph.D. Grant No. PIF 23/246. C.M.-J. acknowledges the predoctoral MICINN grant PRE2019-088519. J. C. acknowledges the Ram\'on y Cajal (RYC2018-025197-I) research fellowship. Authors acknowledge the Quench project that has received funding from the European Union’s Horizon Europe -- The EU Research and Innovation Programme under grant agreement No 101135742, the Spanish Government via the Nanoscale NMR and complex systems project PID2021-126694NB-C21, and the Basque Government grant IT1470-22.


\clearpage

\widetext

\begin{center}
\textbf{ \large Supplemental Material: \\ Robust microwave cavity control for NV ensemble manipulation}
\end{center}

\setcounter{equation}{0} \setcounter{figure}{0} \setcounter{table}{0}\setcounter{section}{0}
\setcounter{page}{1} \makeatletter \global\long\def\theequation{S\arabic{equation}}
 \global\long\def\thefigure{S\arabic{figure}}
 \renewcommand{\theHfigure}{S\arabic{figure}}
 \global\long\def\bibnumfmt#1{[S#1]}
 \global\long\def\citenumfont#1{S#1}
 
 \section{System Hamiltonian}
Polarization transfer protocols consist of a polarizing agent and the target nuclei to polarize. We consider one polarizer (the NV center) and a target nucleus (a $^{13}$C). Acting with a MW driving resonant with the NV leads to the following Hamiltonian:
\begin{equation}
	H = (\omega_n+A_z/2)\sigma_z^n + \omega \sigma_z^e/2 + \frac{1}{2}\sigma_z^e\boldsymbol{A}\cdot\boldsymbol{\sigma}^n + \Omega(t)\sigma_x^e\cos(\omega_dt - \phi).
\end{equation}
where $\sigma_{\{x,y,z\}}^{e (n)}$ are the Pauli spin 1/2 operators for the electron (nucleus). Here, we considered that the state $\ket{-1}$ of the NV is not involved in the dynamics. In a rotating frame with respect to $H_0=\omega\sigma_z^e/2$, and invoking the RWA, we get
\begin{equation}
	H = (\omega_n+A_z/2)\sigma_z^n + \frac{1}{2}\sigma_z^e\boldsymbol{A}\cdot\boldsymbol{\sigma}^n + \frac{\Omega(t)}{2}\sigma_\phi^e + \frac{\delta}{2}\sigma_z^e,
	\label{seq: ham_pulsepol}
\end{equation}
where $\delta=\omega-\omega_d$ and $\sigma_\phi^e=\cos(\phi)\sigma_x^e + \sin(\phi)\sigma_y^e$, while we add a time dependency to the phase ($\phi \rightarrow \phi(t)$) which is equivalent of having drivings with X and Y components. Thus, the pulse Hamiltonian reads (note, for simplicity we remove the $e$ superscript)
\begin{equation}
	H = \frac{\delta}{2}\sigma_z + \frac{\Omega_x(t)}{2}\sigma_x + \frac{\Omega_y(t)}{2}\sigma_y.
	\label{eq: optimization_hamiltonian}
\end{equation}
which is Eq.~(1) of the main text.

\section{\label{supp: cavity} Cavities}
\begin{figure}
	\includegraphics[width=0.5\linewidth]{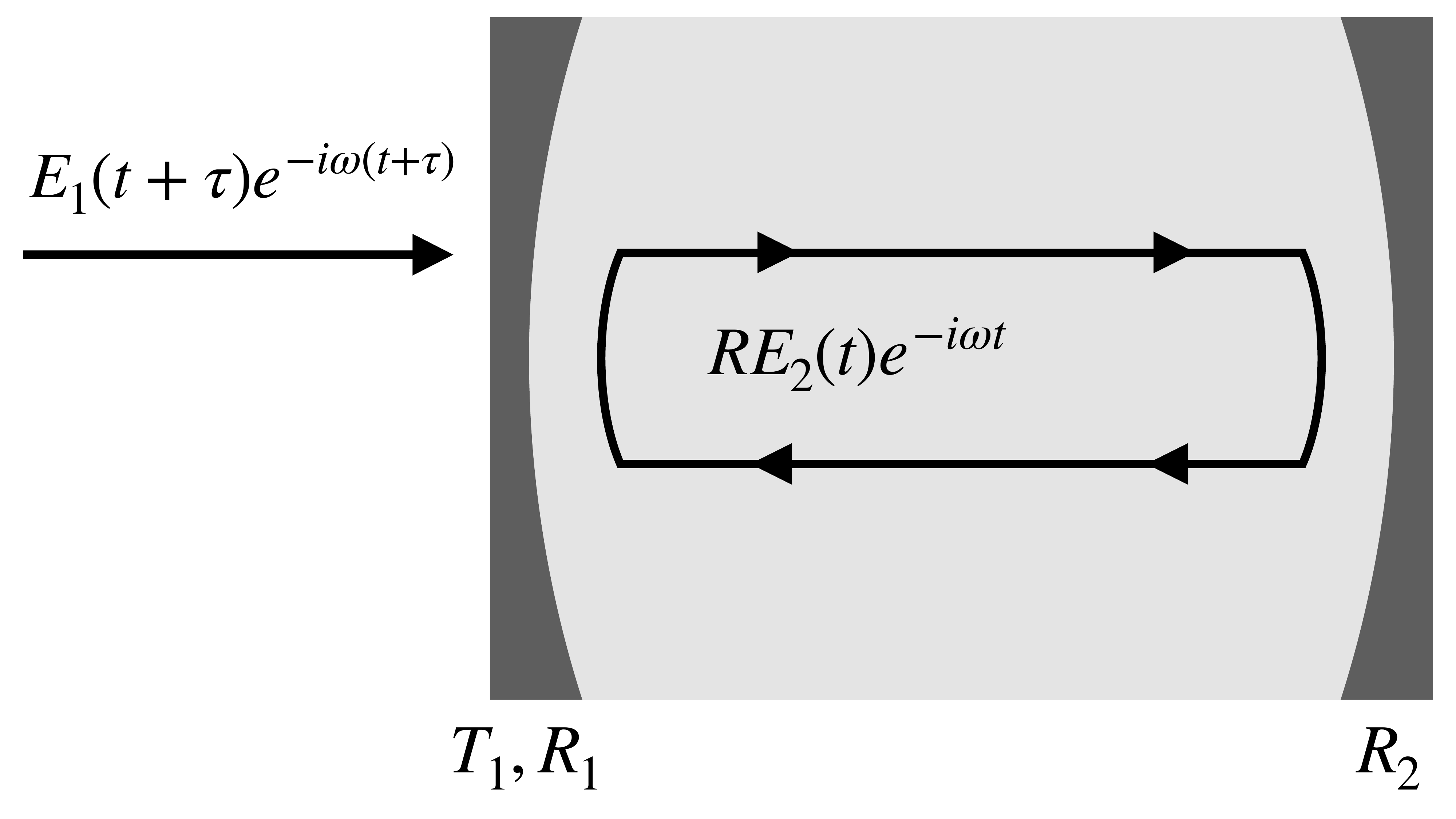}
	\caption{Diagram of fields inside and outside the cavity modeled as a Fabry-Perot interferometer. At every round trip, the external driving will sum to the field inside the cavity.}
	\label{sfig: cav}
\end{figure}

Following Refs.~\cite{DynamicResponse_Lawrence_S, PhaseNoise_Bienfang_S} we model our microwave cavity as a Fabry-Perot interferometer. Such a device consists of two parallel plates, with an external control $E_1(t)e^{-i\omega t}$ and an intra-cavity field $E_2(t)e^{-i\omega t}$ which keeps traveling between the plates (see Fig.~\ref{sfig: cav}). Considering the frequency of the external control $\omega$ is constant, the external control at $t+\tau$ is added to the already present intra-cavity field:
\begin{equation}
	E_{2}(t+\tau)e^{-i\omega (t+\tau)} = \sqrt{T_{1}}E_{1}(t+\tau)e^{-i\omega (t+\tau)} +  RE_{2}(t)e^{-i\omega t}.
	\label{seq: cav_field}
\end{equation}
Here, $R=\sqrt{R_{1}R_{2}}$ is the survival probability of the intra-cavity field after traveling a round trip, see Fig.~\ref{sfig: cav}. Putting the driving on resonance with a frequency of the cavity, the phase accumulated during a round trip is a multiple of $2\pi$, i.e. $\omega \tau = 2\pi k$. Then, Eq.~(\ref{seq: cav_field}) reads
\begin{equation}
	E_{2}(t+\tau) = \sqrt{T_{1}}E_{1}(t+\tau) +  RE_{2}(t).
\end{equation}
Using the two-point finite difference formula to approximate the derivative gives
\begin{equation}
	 \frac{E_2(t+\tau)-E_2(t)}{\tau} = \frac{\sqrt{T_1}}{\tau}\left(\dot{E}_1(t)\tau + E_1(t)\right) + \frac{R-1}{\tau}E_2(t).
\end{equation}
We consider that changes in the external control during a round trip are negligible in comparison to its amplitude. More specifically, this is $\dot E_1(t)\tau \ll E_1(t)$, such that 
\begin{equation}
	\dot{E}_2(t)=\frac{\sqrt{T_1}}{\tau}E_1(t) - \frac{1-R}{\tau}E_2(t).
	\label{seq: e_ode}
\end{equation}
We define the ringing factor: $\gamma=(1-R)/\tau$. We also redefine $E_1$ and $E_2$ as
\begin{equation}
	\begin{aligned}
		E_{1}(t) &= E_\mathrm{max}\cdot f(t)e^{i\phi_{1}(t)} = E_\mathrm{max}\cdot (f_{x}(t) + if_{y}(t)), \\
		E_{2}(t) &= \Omega(t)e^{i\phi_{2}(t)} = \Omega_{x}(t) + i \Omega_{y}(t),
	\end{aligned}
	\label{seq: phase_xy}
\end{equation}
where $E_\mathrm{max}$ is the maximum amplitude of the external control, such that $f(t)$ is normalized
\begin{equation}
	|f(t)|^2=f_x(t)^2+f_y(t)^2\leq 1.
	\label{seq: normalize_f}
\end{equation}

Examining Eq.~(\ref{seq: e_ode}) and setting the external control to its maximum value ($E_\mathrm{max}$), we observe that $\dot{E}_2(t)$ gradually diminishes as $E_2$ asymptotically approaches its maximum value
\begin{equation}
	E_2(t)=\Omega_\mathrm{max} = \frac{\sqrt{T_1}}{1-R} E_\mathrm{max}. 
	\label{seq: omega_max}
\end{equation}

Now we substitute these definitions in Eq.~(\ref{seq: e_ode}), and separate the real and imaginary parts which leads to \cite{PhaseNoise_Bienfang_S}:
\begin{equation}
	\begin{aligned}
		\dot\Omega(t) + \gamma \Omega(t) &= \gamma \Omega_\mathrm{max} f(t) \cos(\phi_2(t)-\phi_1(t)), \\
		\dot{\phi}_2 &= -\gamma \Omega_\mathrm{max} \frac{f(t)}{\Omega(t)}\sin(\phi_2(t)-\phi_1(t)).
	\end{aligned}
	\label{seq: ODE_fp}
\end{equation}
To avoid the singularity when $\Omega(t)=0$, we obtain the equivalent system of equations for the real and imaginary parts of the fields, instead of the amplitude and phase:
\begin{equation}
	\dot\Omega_k(t) + \gamma \Omega_k(t) = \gamma \Omega_\mathrm{max} f_k(t),
\label{seq: ODE_xy}
\end{equation}
with $k\in \{x,y\}$. These equations are linear and analytically solvable for piecewise constant $f_x(t)$ and $f_y(t)$.

\subsection{Analytical solution of the ODE}
The analytical solution of the system in Eq.~(\ref{seq: ODE_xy}) can be obtained treating each of them independently and using an integrator factor $\mu(x,y)$ that transforms the equation $P(x,y)dx + Q(x,y)dy = 0$ in the exact equation
\begin{equation}
    \mu(x,y)[P(x,y)dx + Q(x,y)dy] = 0,
\end{equation}
where $\displaystyle\frac{\partial \mu P}{\partial y}=\frac{\partial \mu Q}{\partial x}$. When the structure of the ODE follows
\begin{equation}
    y' + A(x)y = B(x) \quad \Longrightarrow \quad \left\{
    \begin{aligned}
        P &= (A(x)y-B(x)) \\
        Q &= 1
    \end{aligned}
    \right. ,
\end{equation}
one can prove that the integration factor has the form
\begin{equation}
    \mu(x)=e^{\int Adx}.
\end{equation}
And the equation becomes
\begin{equation}
    e^{\int Adx}y' + e^{\int Adx}A(x)y = e^{\int Adx}B(x) \; \rightarrow \; \frac{d}{dx}\left(e^{\int Adx}y\right)=Be^{\int Adx},
\end{equation}
which results in 
\begin{equation}
    y=e^{-\int Adx}\left(C+\int Be^{\int Adx} dx^\prime\right).
\end{equation}

By substituting $A=\gamma$ and $B= \gamma \Omega_\text{max} f_k(t)$ we get
\begin{equation}
    \Omega_k(t) = e^{-\gamma (t-t_0)} \Omega_k(t_0) + e^{-\gamma (t-t_0)} \Omega_\text{max}\int_{t_0}^{t}\gamma f_k(t^\prime)e^{\gamma (t^\prime-t_0)} dt^\prime,
    \label{seq: ODE_solution}
\end{equation}
which is Eq.~(3) in the main text.

\subsection{Ringing}

\JC{Considering an external constant control $f_k=f_0$, (\ref{seq: ODE_solution}) integrates to
\begin{equation}
	\Omega_k(t) = e^{-\gamma (t-t_0)} \Omega_k(t_0) + \Omega_\text{max} f_0 \left(1-e^{-\gamma (t-t_0)}\right).
	\label{seq: ODE_solution_constant}
\end{equation}
There, we notice an asymptotic behavior of the intra-cavity field towards $\Omega_\mathrm{max}f_0$. To analyze the time scale of this effect, we calculate how long it takes for the intra-cavity amplitude to reach $\Omega_\mathrm{max}/2$ depending on the ringing factor $\gamma$:
\begin{equation}
	\Omega_\mathrm{max}/2 = e^{-\gamma t}\Omega_\mathrm{max} f_0 \left(e^{\gamma t}-1\right)=\Omega_\mathrm{max} f_0 \left(1-e^{-\gamma t}\right) \Longrightarrow f_0-1/2 =  f_0 e^{-\gamma t} \Longrightarrow t = \frac{1}{\gamma} \ln \frac{2f_0}{2f_0-1}.
\end{equation}}
\JC{Note that, in the previous equation we assumed $t_0=0$. However, the general scenario results in 
\begin{equation}
t-t_0 = \frac{1}{\gamma} \ln \frac{2f_0}{2f_0-1}
\end{equation}}

\JC{Eq.~(\ref{seq: ODE_solution_constant}) shows that if $0\ll \gamma (t-t_0)$, leads to $\Omega_k(t)\simeq\Omega_\mathrm{max} f_0$ (note that a large value for $\gamma$ facilitates the condition $0\ll \gamma (t-t_0)$).  This is equivalent to stating that for large ringing factors, intra-cavity amplitudes are proportional to the external controls (as stated in the third paragraph of the introduction).}

\begin{figure}[h]
\centering
	\includegraphics[width=0.8\linewidth]{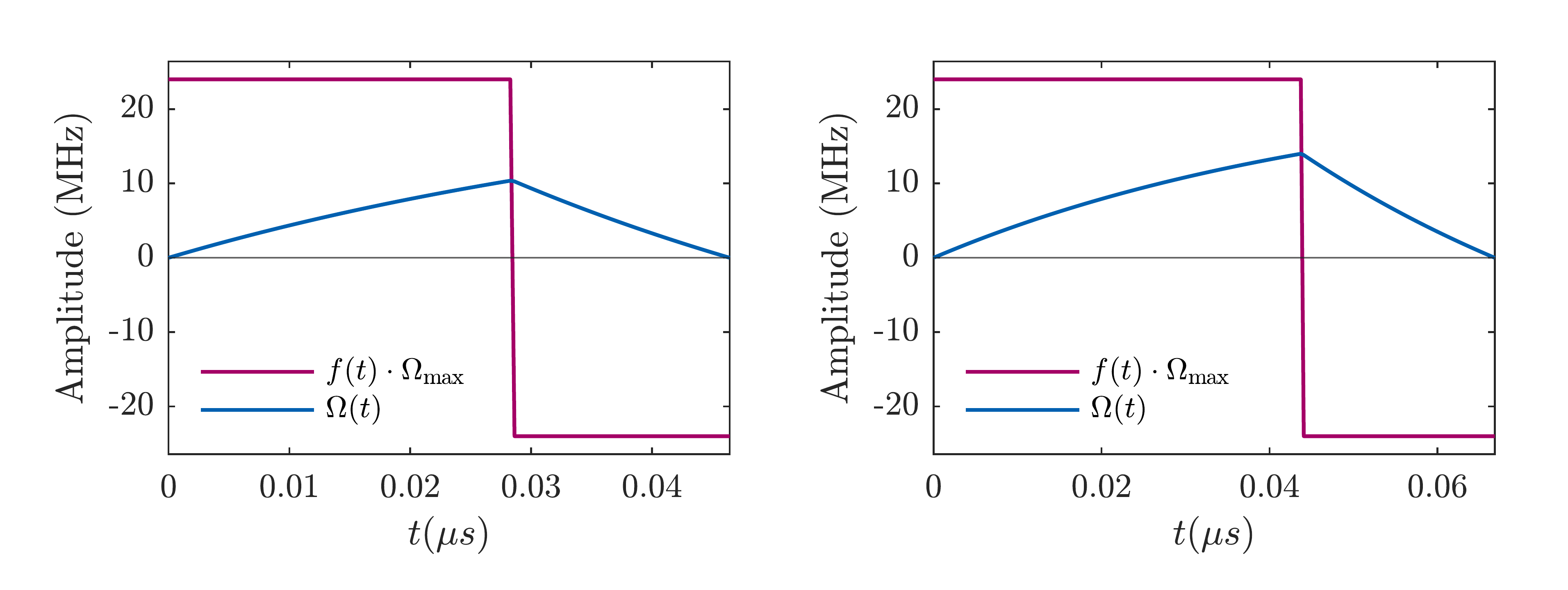}
	\caption{The plot on the left (right) depicts the fastest (and standard) way of applying a $\pi/2$ ($\pi$) pulse on a system inside a cavity. The external control is plotted in pink, and the intra-cavity amplitude in blue. The solution consists of applying an external control with duration $t_1$, followed by a negative, i.e. $\pi$-dephased, control with duration $t_2$. Here we set $\Omega_\mathrm{max}=24\,$MHz and $\gamma=20\,$MHz.}
	\label{sfig: fast_pulse}
\end{figure}

For cavities with small $\gamma$, when the driving is switched off the intra-cavity amplitude decays exponentially, which further evolves the NVs. This effect is called cavity ringing, and in~\cite{Thesis_Tratzmiller_S} an analytical solution is proposed (see Fig.~\ref{sfig: fast_pulse}), so that the intra-cavity amplitude vanishes at the end of the pulse.

\JC{
\subsection{Control algorithms}
Optimal control algorithms are designed to provide controls that generate a desired dynamics in the system. That is, they maximize the fidelity $|\langle U|U_F\rangle|^2$ between a target unitary ($U_F$) and the unitary ($U$) generated by the controls. In cavities with a large ringing factor, $\Omega_x(t)$ and $\Omega_y(t)$ mimic the external controls since their response time is immediate. This is what usual optimal control algorithms, like the well-known GRAPE algorithm~\cite{GRAPE_Khaneja_S}, consider; thus, directly optimizing $\Omega_x(t)$ and $\Omega_y(t)$ is enough. For that, GRAPE provides an analytical expression for the gradient of the fidelity, with respect to $\Omega_x(t)$ and $\Omega_y(t)$ (note the gradient of the fidelity is the central quantity of any gradient-based optimizer).
\begin{equation}
	\frac{\partial}{\partial \Omega_k(t)}|\braket{U|U_F}|^2.
\end{equation}

Meanwhile, in cavities with small ringing factors, the controls suffer non-negligible distortions and specific algorithms that adjust to these effects are required. If the distortion is small, as is the case for intermediate values of $\gamma$, one can employ~\cite{piecewise_Rasulov_S}. There, the authors propose a method that provides controls resilient to small effects generated by transfer functions. However, to obtain accurate controls for arbitrary ringings we designed  an algorithm that considers the dynamics introduced by the cavity in the fidelity $|\langle U|U_F\rangle|^2$. This is, our protocol operates with 
\begin{equation}
	\frac{\partial}{\partial f_k(t)}|\braket{U|U_F}|^2,
\end{equation}
which is the gradient of the fidelity with respect to the external controls $f_x(t)$ and $f_y(t)$.}

\subsection{Derivation of the gradient}
\begin{figure}[h]
	\includegraphics[width=\textwidth]{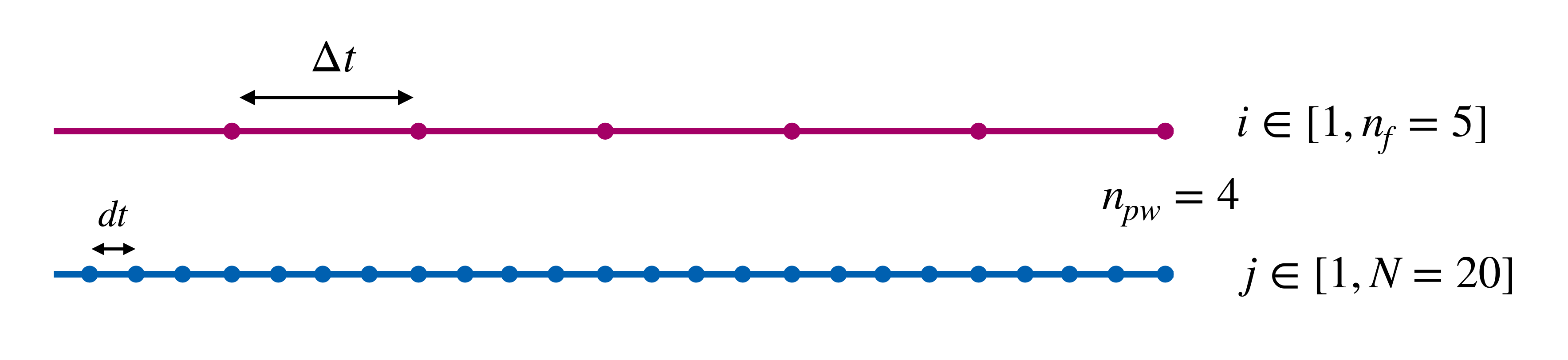}
	\caption{Diagram of the discretization of the external controls (pink) and intra-cavity amplitudes (blue).}
	\label{sfig: discretization}
\end{figure}
We emphasize that the number of steps of $\Omega(t)$, denoted as $N$, has to be large, whereas the one of $f(t)$ ($n_f$) should be kept small in order to have fewer parameters to optimize. Thus, we define $r=N/n_f$ to be an integer number (see Fig.~\ref{sfig: discretization}). Using the following notation
\begin{equation}
	\begin{aligned}
		\lfloor x \rfloor & = \mathtt{floor}(x) = \max\{n\in\mathbb{Z} : n\leq x\} \\
		\lceil x \rceil & = \mathtt{ceil}(x) = \min\{n\in\mathbb{Z} : x\leq n\},
	\end{aligned}
\end{equation}
we rewrite Eq.~(\ref{seq: ODE_solution}) as a sum of integrals (considering $f_k^i=f_k(i\Delta t)$ constant) over the completed $\Delta t$ steps, and an extra term that appears when the last step is uncompleted (i.e. $t<i_c\Delta t$):
\begin{equation}
		\Omega_k(t) = e^{-\gamma t}\Omega_\mathrm{max}\gamma \left( \sum_{i=1}^{i_f}\int_{(i-1)\Delta t}^{i\Delta t} f_k^ie^{\gamma t^\prime}dt^\prime + \int_{i_f\Delta t}^{jdt} f_k^{i_c}e^{\gamma t^\prime} dt^\prime \right),
\end{equation}
where $i_f = \lfloor\frac{t}{\Delta t}\rfloor$ and $i_c = \lceil\frac{t}{\Delta t}\rceil$. And solving the integrals,
\begin{equation}
		\Omega_k(t) = e^{-\gamma t}\Omega_\mathrm{max} \left( \sum_{i=1}^{i_f} f_k^i \left( e^{\gamma i\Delta t} - e^{\gamma (i-1)\Delta t} \right) + f_k^{i_c} \left( e^{\gamma t} - e^{\gamma i_f\Delta t} \right) \right).
\end{equation}
Substituting that $t = jdt$ for $j=1,...,N$
\begin{equation}
		\Omega_k^j = e^{-\gamma jdt}\Omega_\mathrm{max} \sum_{i=1}^{i_f} f_k^i \left( e^{\gamma i\Delta t} - e^{\gamma (i-1)\Delta t} \right) + f_k^{i_c} \left( e^{\gamma jdt} - e^{\gamma i_f\Delta t} \right).
	\label{seq: ODE_sol_disc}
\end{equation}

We calculate $d\Omega_k^j/df_k^i$ differentiating Eq.~(\ref{seq: ODE_sol_disc}), and get a conditional expression on $i$ and $j$, since we can redefine $i_f=\lfloor j/r \rfloor$ and $i_c=\lceil j/r \rceil$, such that they depend on $j$
\begin{equation}
	\frac{\partial\Omega_k^j}{\partial f_k^i} = \left\{
	\begin{aligned}
		& e^{-\gamma jdt}\Omega_\mathrm{max} \left( e^{\gamma i\Delta t} - e^{\gamma (i-1)\Delta t} \right) , \;  i	\leq i_f . \\
		& e^{-\gamma jdt}\Omega_\mathrm{max} \left( e^{\gamma jdt} - e^{\gamma i_f\Delta t} \right) , \;  i_f<i=i_c. \\
		& 0 , \; \text{otherwise}.
	\end{aligned}\right.
\end{equation}
For the sake of simplicity, and after numerical validation, we approximate the gradient as
\begin{equation}
	\frac{\partial\Omega_k^j}{\partial f_k^i} = e^{-\gamma jdt}\Omega_\mathrm{max} \left( e^{\gamma i\Delta t} - e^{\gamma (i-1)\Delta t} \right)
\end{equation}
for every $j=1,...,N$ and $i=1,...,i_c$.

\section{Energy constrained pulses}
From Eq.~(\ref{seq: omega_max}) one could think that the maximum intra-cavity amplitude achievable can grow indefinitely, as we increase the amplitude of the external control $E_\mathrm{max}$. However, when increasing the input power the antenna overheats \cite{versatile_Pellicer_S} and the performance worsens, making us fix an affordable value of $\Omega_\mathrm{max}$. To ensure the algorithm provides pulses that obey the constraint of~(\ref{seq: normalize_f}), we verify that $f(t)$ remains within the unit circle for all $t$. In the case it falls outside, we project it onto the boundary (the circumference of radius one) keeping the same angle $\theta = \tan^{-1}\left(\frac{f_y(t)}{f_x(t)}\right)$. We repeat this process every time the optimizer updates the external control.

\section{Pulses along other axes}
\begin{figure}[h]
	\includegraphics[width=0.5\textwidth]{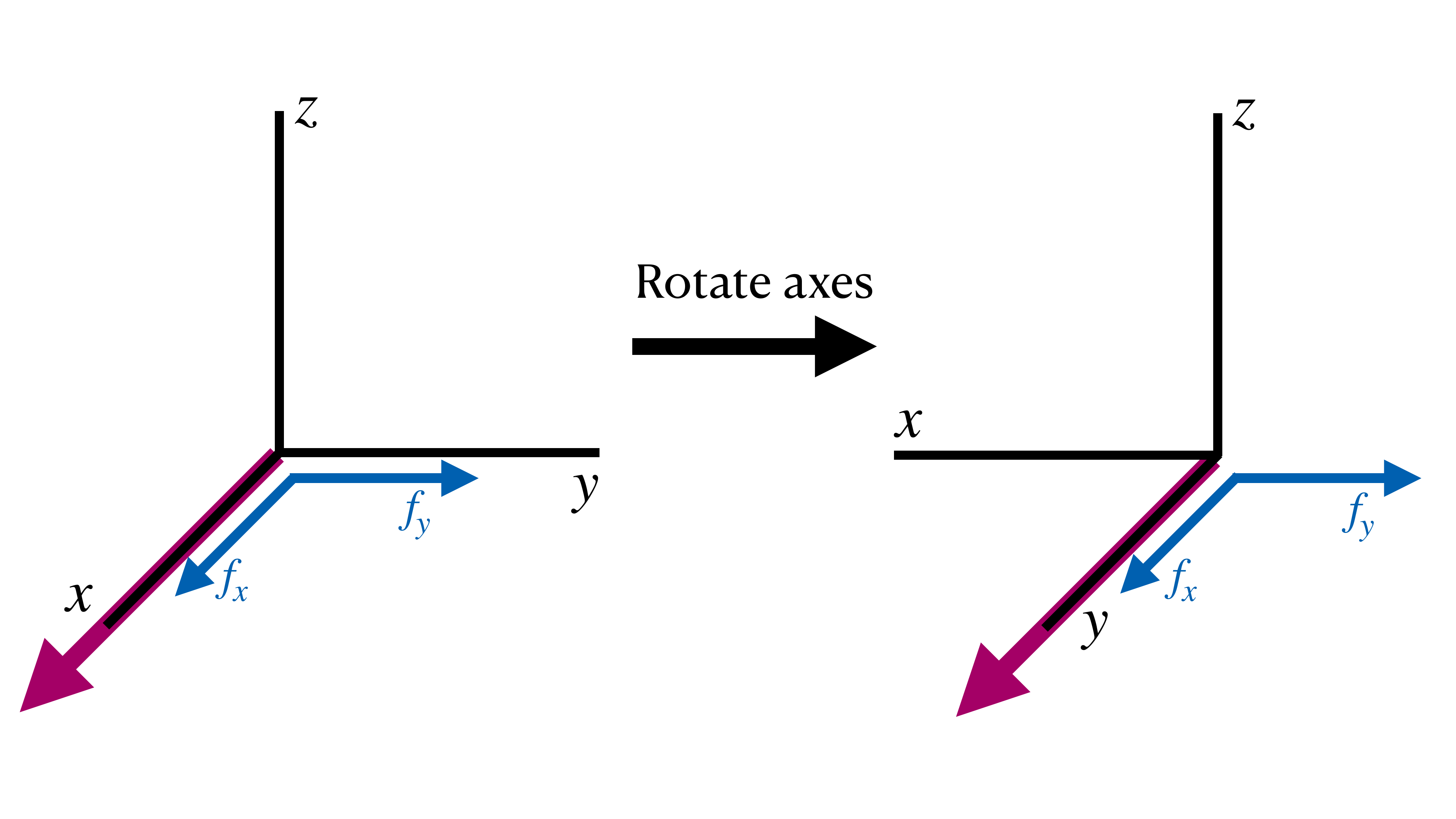}
	\caption{Diagram of the transformation. The pink arrow indicates along what axis happens the rotation. The blue arrows are the controls corresponding to that rotation.}
	\label{sfig: rotate_axes}
\end{figure}

To build the PulsePol sequence, $\pi$ and $\pi/2$ pulses are also required along the Y-axis. While we have optimized the controls for pulses along the X-axis, the controls for the Y-axis pulses can be obtained by applying a specific transformation to the original set. In Fig.~\ref{sfig: rotate_axes} we rotate the axes to make the Y-axis coincide with the pink arrow. Now, the arrow $f_x$ coincides with the Y-axis, and $f_y$ with the -X-axis. So, the transformation is $f_x^\prime=-f_y$ and $f_y^\prime=f_x$; equivalent to $f^\prime_x+if^\prime_y = e^{i\pi/2}fe^{i\phi_1}=-f_y + if_x$, see~(\ref{seq: phase_xy}).

\JC{
\section{Robustness}
To further assess the error robustness of the protocol, we introduce inhomogeneities in the microwave (MW) field across the diamond. These variations in $\Omega_\mathrm{max}$ within the antenna range from 1\% to 4\%~\cite{3D_Kapitanova_S, Circular_Yaroshenko_S}. In the simulations shown in Fig.~\ref{sfig: robustness}, we assume a Gaussian error with 1\% deviation in the external controls $f_k(t)$.

Additionally, we model the spatial dependence of $\Omega_\mathrm{max}$, assuming it decreases as the NV center moves away from the antenna center, following the relation $\Omega_\mathrm{max}\rightarrow \Omega_\mathrm{max}(1-|\epsilon|)$, where $\epsilon\in [-1,1]$ is treated as a constant. To analyze its impact on the polarization curve, we compute a weighted sum over $\epsilon$, with weights following a normalized Gaussian distribution with a standard deviation of 4\%. This accounts for variations of up to 10\% while assigning greater importance to smaller deviations, as a higher density of NV centers is expected in those regions. The chosen deviation corresponds to the worst-case scenario among the referenced values. Indeed, in Refs.~\cite{3D_Kapitanova_S, Circular_Yaroshenko_S} we find maximum errors of 1\% and 4\%.

Ultimately, we reconfirm that the PulsePol sequence with optimized pulses achieves greater polarization transfer compared to the sequence constructed with standard pulses (see Fig.~\ref{sfig: robustness} (c)).

\begin{figure}[h]
	\includegraphics[width=\textwidth]{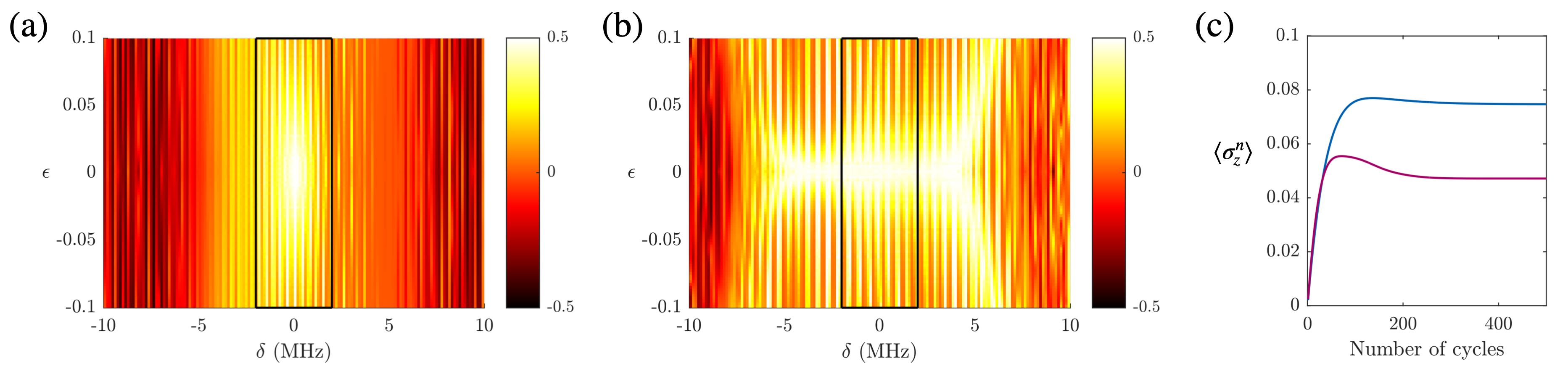}
	\caption{\JC{Comparison of the robustness of the PulsePol with the standard vs optimized pulses. Figures \textbf{(a)} and \textbf{(b)} depict the nuclear polarization $\langle \sigma_z^n \rangle$ (see color-bar) as a function of errors $\delta$ and $\epsilon$ after applying the protocol built with the standard pulses and our optimized pulses, respectively. Two PulsePol sequences were applied before reinitializing the NV completing one cycle, and 500 cycles were performed before measuring the nuclear polarization. In \textbf{(c)} we depict the average polarization curve inside the regions marked by the squares in figures (a) and (b). These regions are delimited by the constraints $|\delta| \leq 2\,$MHz and $|\epsilon| \leq 0.1$. The curve achieved with our optimized external controls (blue) exceeds the one obtained with the standard approach (pink). The values of the parameters used in the simulations are $\gamma=20\,$MHz, $\Omega_\mathrm{max}= 24\,$MHz, $B_z=0.015\,$T,  $A_x=4\,$kHz and $A_z=3.7\,$kHz. Note that, this selection (i.e., the parameter regime from which we harvest polarization) favors the standard approach while penalize ours as it only considers detunings $\leq 2\,$MHz. Thus, in a real scenario where detunings could exceed 2 MHz, our approach would be even more effective for polarization transfer than that shown in (c).}}
	\label{sfig: robustness}
\end{figure}
}

\end{document}